# BARYONIC DARK HALOS: A MODEL WITH MACHOS AND COLD GAS GLOBULES

ORTWIN GERHARD[1] AND JOSEPH SILK[2]
[1]*Astronomisches Institut, Univ. Basel, Switzerland*
[2]*Dpt. of Astronomy, Univ. of California, Berkeley, U.S.A.*

**Abstract.** The dark matter in the halos of galaxies may well be baryonic, and much of the mass within them could be in the form of clusters of substellar objects within which are embedded cold gas globules. Such halos might play an active role in galaxy formation and evolution.

## 1. Introduction: The Case for Baryonic Halos

A number of arguments point towards the dominance of baryonic dark matter in galaxy halos. Some of the strongest arguments have been known for many years: (i) Cosmic nucleosynthesis predicts considerably more baryons than are seen in luminous form (Kolb & Turner 1990). (ii) While the visible mass in spiral galaxies fails to explain the *amplitudes* of the observed rotation curves, their *form* is often well-predicted from scaling up the distribution of gas mass (Bosma 1981). With the discovery of MACHOs, we now have evidence for *some* of the halo dark matter being baryonic (Alcock *et al.* 1993, Aubourg *et al.* 1993). Since this is the only observational evidence to date on the nature of any dark matter, it is natural to ask whether the entire mass in galaxy halos could be baryonic.

## 2. The Old Model: Gas Clouds Only

In this talk a brief account is given of a model of Gerhard & Silk (1995; hereafter GS) which asserts that a substantial fraction of halo dark matter may be in the form of dense, cold gas clouds. In its original form, as submitted to Nature in 1993, the model considered these gas clouds as isolated objects. We found that collisional dissipation and evaporation can be avoided if the clouds have sufficiently high column density and correspondingly low area filling factor, and for the same reasons detection in emission or absorption is very difficult. However, as emphasized by several colleagues, the theoretical problem of preventing the clouds from collapsing and forming stars is more serious than in the disk ISM because of the long time-scales involved. The nearby gas might moreover have been seen in $\gamma$-rays via cosmic ray interactions and ensuing $\pi^0$-production, and in UV-illumination when these clouds traversed a nearby HII-region.

## 3. The New Model: Gas Globules Stabilized by Surrounding MACHO Clusters

In the meantime, MACHOs of mass $\sim 0.01\,\mathrm{M}_\odot$ had been discovered, and been found to contribute an appreciable fraction (Alcock *et al.* 1995) of the Milky Way's halo mass. This implies that at early times some halo gas *did* fragment and collapse, and the question arises as to whether all the remaining gas collapsed onto the disk or whether most of it remained in the halo as dense cold clouds. In fact, since such low mass MACHOs do not inject as much energy into their surroundings as do normal main sequence stars, the gas might remain near the MACHOs. Further, we realized that, if the MACHOs are clustered by analogy with the formation of most low-mass stars, they can exert a stabilizing influence on any remaining embedded gas by removing some of the self-gravity response. We have considered this in the context of a simple model in which both the MACHO cluster and the gas cloud are spherical and do not rotate, and in which the equation of state is polytropic. One then finds that $\sim 10 - 50\%$ of the MACHO cluster mass can be embedded as a stable sphere of gas. For more general configurations, up to of order half the total mass could be stabilized. This result has led to a modified model in which the halo is assumed to be made of clusters of low-mass objects within which cold gas clouds are embedded.

## 4. Physical Parameters of the Gas Clouds

The thermal state of these clouds is uncertain, both because of unknown cosmic ray heating rates in the outer halo, and because modern cooling calculations with revised molecular excitation rates (Neufeld *et al.* 1995) are not available for the low abundances and cosmic ray ionization rates expected in the outer halo. However, at the high densities implied by the model ($n \sim 10^7\,\mathrm{cm}^{-3}$), the cooling rate per molecule appears to decrease with density. If the temperature remains above the CMB value, this will lead to a polytropic rather than an isothermal equation of state.

A critical condition for gas to survive in the halo is that the area filling factor of the clouds is sufficiently small, $f \sim 0.01$ in the outer halo. Given the rotation curve, this sets a lower limit to the cloud column density, $N \sim 10^{23}\,\mathrm{cm}^{-2}$ if collisions are to be avoided in the outer halo, and larger for clouds further in. Together with a temperature of, say, $T \simeq 10\,\mathrm{K}$, the condition that the clouds be partially self-gravitating determines their mass and radius. Typical values are $M_c \lesssim 1\,\mathrm{M}_\odot$ and $L \lesssim 0.02\,\mathrm{pc}$.

## 5. Parameters of the MACHO Clusters

The masses of individual fragments are limited by the opacity argument to $m \gtrsim 10^{-3}\,\mathrm{M}_\odot$ (Rees 1976), and the MACHO experiments favour masses somewhat larger. So that the gas content of the halo is non-negligible, the MACHO cluster mass should not exceed a small multiple of the mass of its embedded gas cloud. The requirement that the cluster should not evaporate in a Hubble time (Moore & Silk 1995) is then marginally satisfied, requiring some fraction of cloud support in macroscopic motions. It also sets an upper limit to the cloud column density. Because the constraints on cloud collisions require high column densities, whereas the constraints on cluster evaporation favour low column densities, the new model now leads to fairly well-defined typical parameters: cluster masses of $\sim 10\,\mathrm{M}_\odot$, MACHO masses of $0.01\,\mathrm{M}_\odot$, and cold gas content up to of order 50%, with typical column density $N \sim 10^{23}\,\mathrm{cm}^{-2}$ and temperature $T \sim 10\,\mathrm{K}$.

## 6. Observational Constraints

With these parameters, one furthermore predicts that the dissipational evolution is a strong function of radius, in the sense that inner halos should by now be largely depleted of gas, while outer halos should still contain substantial gas fractions. This also greatly reduces the observational problems that remained in the original model. While at present, observations cannot constrain our hypothesis, nevertheless a number of observational techniques can be improved to search for the postulated cloud-clusters. In particular, more sensitive FIR, gamma ray, local and high redshift mm-line observations, and further analysis of the microlensing experiments will provide tighter constraints on the cloud parameters and might falsify the model.

## 7. Implications

If indeed a significant fraction of the mass of galactic halos is or once was in the form of cold gas, then the radial dependence of the cloud survival rate has cosmogonical implications for galaxy formation. The halos of galaxies would then play a much more active role in the build-up of galactic disks and their chemical evolution. Moreover, these processes would then naturally occur fastest in massive galaxies, with much of the mass in dwarf systems remaining in dark halos even today.

## Discussion

RIX: If you have 0.01 of sky covered with dense, presumably optically opaque, molecular clouds, would you not expect a distinct time-dependent absorption signature in the MACHO and OGLE surveys? If so, what is the time-scale for such time variations?

GERHARD: The time-scale is $\sim 80$ yr for our standard parameters, so it should be detectable statistically. We have not worked the details out yet.

MOORE: The dynamics of the Magellanic Stream (Moore & Davis 1994) constrain the total mass of diffuse gas within 50 kpc to be $\sim 0.5\%$ of the total halo mass - consistent with ROSAT observations. Is it possible to confine 99.5% of the halo mass in such small cold gas clouds, without winds, evaporation etc. liberating a significant fraction of gas to a diffuse component?

GERHARD: The evaporation rate is about $10^{-11}$ yr$^{-1}$ for a hot gas density of $10^{-4}$ cm$^{-3}$; collisions yield a comparable rate of $\gtrsim (100 t_{\rm dyn})^{-1}$. This is about consistent with 1% of the halo gas in diffuse form if the relevant cooling time is $10^9$ yr.

BOSMA: There are galaxies which have flat rotation curves, yet with a light distribution completely different from a standard exponential disk (Malin-1 "cousins"). If the dark halo participates in the evolution of the stellar populations, how do you explain such large differences between such galaxies and the more "standard" ones?

GERHARD: We have not tried sofar, but these halos could be unusually diffuse or they could have formed late.